\documentclass[twocolumn,showpacs,preprintnumbers,amsmath,amssymb]{revtex4}
\usepackage{graphicx}
\usepackage{dcolumn}
\usepackage{bm}

\begin{document}

\title{Electronic Structure of Nearly Ferromagnetic compound HfZn$_{2}$ }

\author{T.Jeong}
 
\affiliation{
Department of Physics, University of California, Davis, California 95616
}


\begin{abstract}
The electronic structure of HfZn$_{2}$ has been 
studied based on the density functional theory within the 
local-density approximation.
The calculation indicates that HfZn$_{2}$ shows
ferromagnetic instability. 
Large enhancement of the static 
susceptibility over its non-interacting value is found due to 
a peak in the density of states at the Fermi level.

\end{abstract}


\maketitle

\section{\label{sec:level1}Introduction}

The discovery of ferromagnetic superconducting phase of ZrZn$_{2}$ 
has been revived both theoretical and experimental attention\cite{pfleiderer}.
Investigation the 5d compound HfZn$_2$ which has the same 
crystal structure (C15 cubic Laves structure)
with the ZrZn$_{2}$ should be quite interesting.
The lattice constant of HfZn$_{2}$(a=7.32$\AA$) is very similar 
to that of ZrZn$_{2}$(a=7.39 $\AA$).
The isoelectronic isostructural material 
ZrZn$_{2}$ is nowadays considered 
a classic example of a Stoner-Wohlfarth itinerant weak ferromagnet.
The magnetic moments of ZrZn$_{2}$ have been reported as 
very small magnetic moments(values from 
0.12 to 0.23 $\mu_{B}$ )\cite{ZrZn1, ZrZn2}. 
These do not saturate 
even at magnetic fields as high as 35T.
The curie temperature T$_{c}$ drops approximately linearly
with pressure, starting at 29K at $P=0$ and 
decreasing to 4K at $P=16$ kbar, 
which extrapolates to a quantum critical point(QCP) at 
$P=18-20$ kbar. 
We contend that the weak ferromagnetism of ZrZn$_{2}$ is an accidental 
consequence of its band structure: apparently $E_{F}$ is located 
near a sharp maximum in the density of states. If this picture is correct, 
then it should follow that the analogous $5d$ compound HfZn$_{2}$  
which has the same crystal structure, and same number of $d$-electrons, 
should behave similarly.

Experiments show that HfZn$_{2}$ is an exchange enhanced 
paramagnet \cite{knapp}.
Though the material does not manage to reach the ferromagnetic 
state, its susceptibility $\chi(T=0)$ is uniquely high among 5d systems and 
strongly temperature dependent; in fact,
it is about the same as the susceptibility of the nearly 
ferromagnetic 4d element Pd ($\chi=6.8\times 10^{-4} $emu/mol)\cite{Pd}.
The measured susceptibility follows a Curie-Weiss law in the
temperature range of 2K-294K. 
Such temperature dependence is found in rare-earth and some 3d 
elements, but it is unusual for a 4d and 5d compound.
No other compound exhibits a Curie-Weiss susceptibility at 
low temperature but only 4d compound ZrZn$_{2}$ shows 
the temperature dependence.
The measured susceptibility data can be expressed as 
$\chi(T)=\chi_{0}+\frac{C}{T-\theta}$, and Knapp {\it et al} \cite{knapp}
obtained $\theta =-160$K and $\chi_{0}=167\times 10^{-6}$ emu/mole.
The negative Curie-Weiss $\theta$ does not result from antiferromagnetic 
order by their experiment.
Such negative $\theta$ has been observed in the Kondo systems, which 
are dilute alloys of 3d transition metals in nonmagnetic hosts.
In such systems the negative $\theta$ are taken as an indication 
of strong interactions between the local 3d moments and the  
conduction electrons of the hosts.
When the negative $\theta$ are observed in an alloy, the 
Kondo resistivity minimum is also found.
If HfZn$_{2}$ is related with the local moments like 
the dilute system, we can expect that HfZn$_{2}$ has 
a Kondo type resistence.
Experiments show that 
the resistivity data follows $\rho=\rho_{0}+AT^{3}$
for the temperature range of T $\leq $ 40K, but no resistivity minimum
occured\cite{knapp}.
They also measured the linear specific heat
coefficient for HfZn$_2$ of $\gamma$=15.8 mJ/K$^{2}$ mole(formular unit).  
The large zero temperature magnetic susceptibility is much larger 
than that expected from the measured electronic specific 
heat coefficient $\gamma$, which indicates that exchange enhancement 
effects are sufficient enough to make the system ferromagnetic.

In this work,
the precise self-consistent full 
potential 
local orbital minimum basis band structure scheme (FPLO) are 
employed to investigate the electronic and magnetic 
properties of HfZn$_{2}$ based on the density functional theory.
We compare the electronic structures and magnetic properties  
between HfZn$_{2}$ and ZrZn$_{2}$.
We focus on studying the effect of magnetism on the band structure, Fermi 
surfaces and compare them with the experimental results.

\section{Crystal Structure}

HfZn$_{2}$ has same crystal structure with ZrZn$_{2}$. 
They crystallize into a C15 cubic Laves lattice.
The C15 (AB$_{2}$ ) structure is a closely packed structure 
and the site symmetry is high for the two constituents. 
In this structure, Hf atoms occupy the positions of a 
diamond lattice while the Zn atoms form a network of interconnected 
tetrahedra. 
Since the major contributions to $N(E_{F})$ come from 
Hf, the local environment of Hf atoms is particulary important 
for our concerns.
Each Hf is surrounded by Zn  neighbours at a distance of 
2.70  $\AA$
and Hf neighbours 2.82 $\AA$ away. 
HfZn$_2$ structure belongs to the Fd3m
space group with Hf occupying the 8a site, and Zn the 16d site.
The site symmetry of Hf is $\bar{4}3m$ and Zn has $\bar{3}m$ site 
symmetry.
We used experimental lattice constant (7.32$\AA$) for our calculations.
There are 2 formular units per cell.
 
\begin{figure}
\includegraphics[height=8.5cm,width=8.5cm,angle=-90]{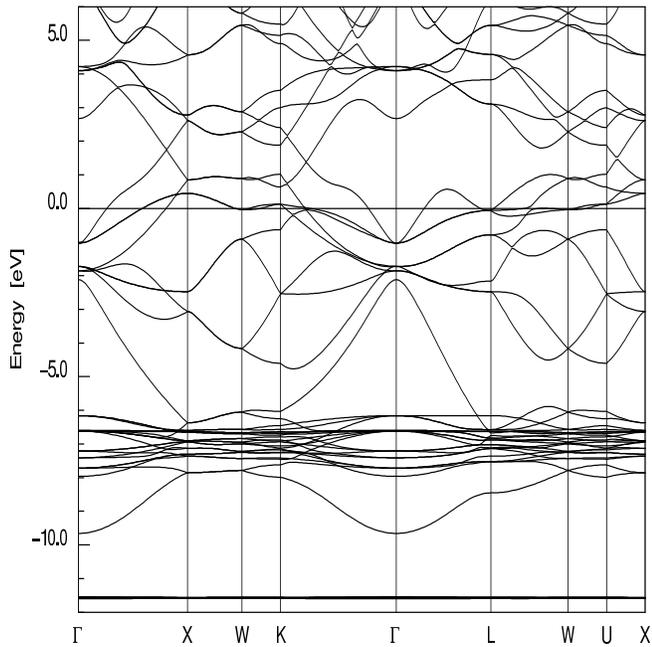}
\caption{The full LDA bandstructures of non-magnetic HfZn$_{2}$ along
symmetry lines, showing that there are several bands with dispersion 
beging of primarily Hf $5d$ and Zn $4p$ characters near the Fermi level.} 
\label{fullband}
\end{figure}

\begin{figure}
\vskip 5mm
\includegraphics[height=8.5cm,width=8.5cm,angle=-90]{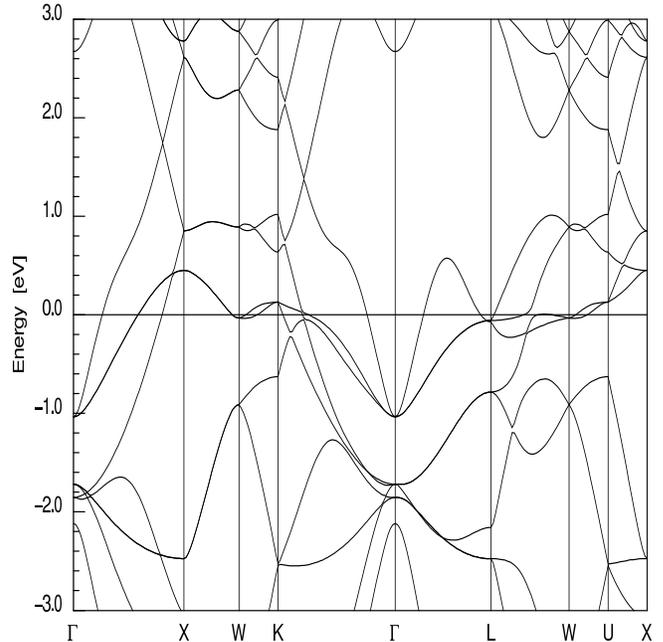}
\vskip 5mm
\includegraphics[height=8.5cm,width=8.5cm,angle=-90]{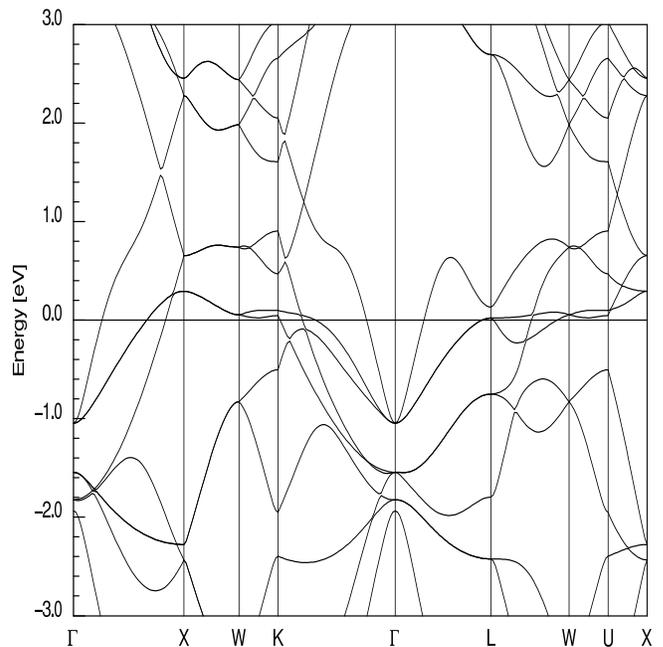}
\caption{Top panel: The LDA non-magnetic band structure of  
HfZn$_{2}$ showing that the Hf $5d$ states play dominant roles 
near the Fermi level.
Bottom panel: Band structure of ZrZn$_{2}$; the Zr 4d states are 
important near the Fermi level.} 
\label{band2}
\end{figure}

\section{Method of Calculations}

We have applied the full-potential 
nonorthogonal local-orbital minimum-basis (FPLO) scheme within the local 
density approximation (LDA).\cite{koepernik}
In these scalar relativistic calculations we 
used the exchange and correlation potential of Perdew and Wang.\cite{perdew}
Hf $4s, 4p, 4d, 4f, 5s, 5p$ states and Zn $4s,4p$ were included as 
valence states. All lower states were treated as core states.
The inclusion of the relatively extended semicore states of
Hf $4s,4p,4d, 4f,5s, 5p$ and Zn $4s, 4p$ 
as band states was done because of the considerable overlap of these 
states on nearest neighbors.
This overlap would be otherwise neglected in our FPLO scheme. 
The spatial extension of the 
basis orbitals, controlled by a confining potential $(r/r_{0})^4$, was 
optimized to minimize the total energy. The self-consistent potentials were 
carried out on a k mesh of 24 k points in each direction of the Brillouin zone,
which 
corresponds to 413 k points in the irreducible zone.
A careful sampling of the Brillouin zone is necessitated by the 
fine structures in the density of states near Fermi level E$_{F}$.

\section{Results}

\begin{figure}
\vskip 5mm
\includegraphics[height=8.5cm,width=8.5cm,angle=-90]{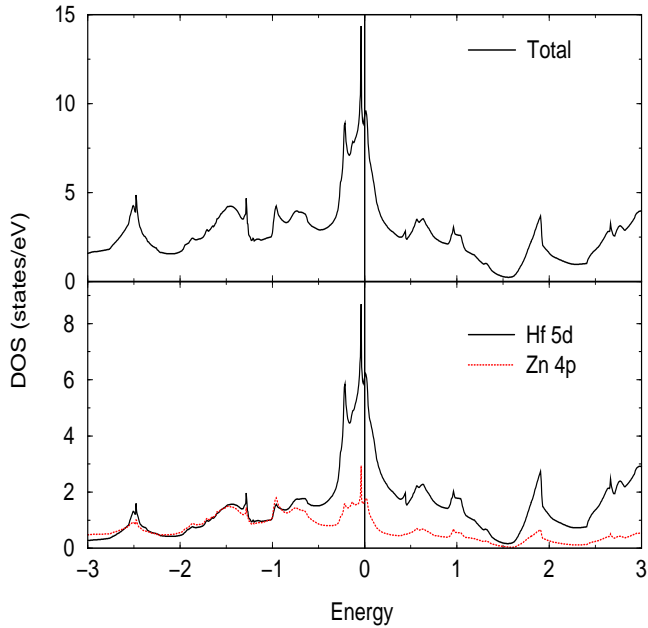}
\caption{The projected density of states of HfZn$_{2}$. 
The contribution of Hf $5d$ and Zn $4p$ to the density of states } 
\label{dos}
\end{figure}

We first show the full LDA non-magnetic
band structure of HfZn$_{2}$ in the Fig. \ref{fullband}.
The very flat Hf $4f$ states lie at -11.5 eV.
Another flat Zn $3d$  bands are located  
between -8.0 eV and -6.0 eV. 
Above them there are Zn $4s$ and Hf $6s$ bands.
Those bands near Fermi level are mainly Hf-centered 
$5d$ states with hybridization of Zn $4s$. 
The Zn $4s$ states cross the Fermi level at the symmetry point 
$\Gamma$ with hybridization of Hf $dxy,dyz$ and $dxz$ states, which make
the ball-like Fermi surface(FS).
Here we compare the band stuructures between HfZn$_{2}$ and ZrZn$_{2}$ with 
more details in the Fig.\ref{band2}. 
While the Zr $4d$ states are important near the
Fermi level ($E_{F}$) in the ZrZn$_{2}$ compound,  
the Hf $5d$ states play dominant roles near the $E_{F}$ in HfZn$_{2}$.
The Zn $3d$ states are located around -7.0 eV in both ZrZn$_{2}$ and 
HfZn$_{2}$.
Density of states (DOS) is shown in Fig. \ref{dos}.
The Fermi energy falls extremely close to the edge od a very 
narrow peak in the DOS. The DOS peak arises from Hf $5d$ bands 
hybridized with Zn $4s$ states.
The bandstructure provides explanation of the high
peaks in the density of state, 
in particular close to the Fermi level.
While the bands in $\Gamma$KW plane exhibit sizable dispersion
their counterparts
in the LWU plane are rather flat. 

Important insight into magnetic fluctuation effects is provided by 
the enhancement of the bare Pauli susceptibility in a metal. We have
evaluated, within the density functional theory formalism, the Stoner 
enhancement of the susceptibility 

\begin{eqnarray}
\chi=\frac{\chi_{\circ}}{(1-IN(E_{F}))}\equiv S\chi_{\circ} ,
\label{susceptibility}
\end{eqnarray}

where $\chi_{\circ}=2\mu_{B}^{2}N(E_{F})$ 
is the non-interacting susceptibility 
and $S$ gives the electron-electron enhancement in terms of the Stoner 
constant $I$. We have calculated $I$ using both the Janak-Vosko-Perdew
theory \cite{jan77}
and fixed spin moment calculations. 
We obtain $N(E_{F})I=1.1$
which indicates ferromagnetic instability.
With the calculated value of $N(E_{F})$=4.65 states/eV for HfZn$_2$
, we get the Stoner $I$ of 0.24.  
The presence of a peak close below the Fermi level
suggest that a very small hole or electron doping can drive system into
ferromagnetic regime.
In our calculation the spin-orbit coupling was neglected.
In ZrZn$_{2}$ case the $N(E_{F})I=1.022$ which corresponds to 
the magnetic instability.
Knapp {\it et al.} also measured the linear specific heat
coefficient for HfZn$_2$ of $\gamma$=15.8 mJ/K$^{2}$ mole(formular unit). 
The calculated value of $N(E_{F})$=4.65 states/eV for HfZn$_2$
corresponds to
a bare value $\gamma_b$=10.88 mJ/K$^{2}$ mole(formular unit), 
which is smaller than the experimental one.
In contrast the linear specific heat
coefficient for ZrZn$_2$ is relatively high of the 
value $\gamma$=47 mJ/K$^{2}$ mole(formular unit).

\section{Discussion}

Motivated by the observation of Curie-Weiss susceptibility in 
weakly ferromagnetic metals such ZrZr$_{2}$ and Sc$_{3}$In where 
the local moment picture is clearly inadequate,
Moriya {\it et al.}\cite{moriya, ueda} developed a theory called 
self consistent renormalization (SCR) theory of ferromagnetic metals. 
A self-consistent treatment needs to calculate dynamical 
susceptibility $\chi(q, \omega)$ and free energy  
so that the static long wavelength limit of the dynamical susceptibility 
agrees with that calculated from the renormalized free energy. 
Ogawa \cite{ogawa} did an experiment with the  
Zr$_{1-x}$Hf$_{x}$Zr$_{2}$, which 
is consistent with the SCR features. 

Moriya and Kawabata \cite{moriya} discussed the importance 
of the effects of spin fluctuations on magnetic 
properties of weak itinerant ferromagnets.
Spin fluctuations affect also an electrical conductivity of 
weak itinerant ferromagnets. It gives a large enhancement of 
the electrical resistivity at low temperature.
Mills and Lederer \cite{mills} found out that 
 T$^{2}$ dependence of resistivity is a general feature 
of a Fermi liquid in a low temperature range. 
Mathon \cite{mathon}discussed the temperature
dependence of resistivity of weak ferromagnets near their 
critical concentration, giving a result of T$^{5/3}$ dependence. 
Ueda and Moriya \cite{ueda} developed a theory 
of electrical and thermal resistivities on the basis 
of Moriya and Kawabata's studies, which treated the spin 
fluctuation effect in a self-consistent way and 
has given a good explanation of the magnetic properties observed 
in weakly and nearly ferromagnetic metals.
They give expressions to the resistivity of weak ferromagnets 
not only in low and high temperature limits but also in intermediate 
temperature region including T$_{c}$. 
Their theory clearly shows that
the spin fluctuation 
effect is demonstrated not only in the magnetic properties but also
in the electrical conduction of weak itinerant ferromagnets.
According to the experiment by Knapp {\it et al.} the resistivity data of 
HfZn$_{2}$ follows the formular $\rho=\rho_{0}+AT^{3}$
for the temperature range of T $\leq $ 40K where $A=1.1 \times 
10^{-4}\mu\Omega$ cm/deg and $\rho_{0}=36\mu\Omega$ cm.
The $T^{3}$ term is indicative of strong $s-d$ scatter, and this
result does not follow the theory of Moriya {\it et al.}\cite{moriya}.

Even though we can learn the some physical properties of HfZn$_{2}$ from 
the results of Ogawa and Knapp's experiments, further studies are required
in both experiment and theory.   
For example the Curie-Weiss temperature dependence of $\chi$ of 
ZrZn$_{2}$ and HfZn${2}$ shows 
the characteristics of localized moments behaviors. 
But the large $T^{3}$ 
in the resistivity and magnetic behavior ZrZn$_{2}$ below $T_{c}$ 
are indicative of the itinerant model.
Therefore the both model should be incorporated to explain 
this phenomena.

\section{Conclusions}

In this article we presented the electronic band structures 
of HfZn$_{2}$ and compared them with the ZrZn$_{2}$.
In the band structure of HfZn$_{2}$, Hf 5d states play 
dominant roles near the Fermi level while Zr 4d states are 
important in the ZrZn$_{2}$.
The presence of the peak close the Fermi level in the Density of states
of HfZn$_{2}$ suggests that a very small hole or electron doping
can drive system into ferromagnetic regime.  
The experiments show
that  HfZn$_{2}$ is located at
the boundary between non-magnetic and ferromagnetic ground state.
Our fixed spin moment calculations show that HfZn$_{2}$ has 
the magnetic instability, which is in agreement with the experiment.
Since HfZn$_{2}$ is near the boundary of a non-magnetic 
and ferromagnetic state, the experiment for the phase 
transition from paramagnet to ferromagnet would be very interesting. 
Also the experimental investigation of the resistance behavior 
in the broad range of temperatures should be quite interesting.


\end{document}